\documentclass[noshowpacs,pre,twocolumn,preprintnumbers,superscriptaddress,amsmath,amssymb,floatfix]{revtex4}
\usepackage{amsmath}
\usepackage[caption=false]{subfig}
\usepackage{graphicx}
\usepackage{color}
\usepackage{placeins}
\usepackage{tikz}
\usepackage{mathtools}
\usepackage{amsthm}
\usepackage{xkeyval,xcolor}
\usepackage{float}

\newcommand{\er}{Erd\H{o}s-R\'{e}nyi}
\newcommand{\av}[1]{\langle #1 \rangle}

\newcommand{\RN}[1]{%
	\textup{\uppercase\expandafter{\romannumeral#1}}%
}

\begin{document}
%opening
\title{Distance Distribution in Extreme Modular Networks}

\author{Eitan Asher}
\thanks{These two authors contributed equally}
\affiliation{Department of Physics, Bar-Ilan University, Ramat Gan, Israel}
\author{Hillel Sanhedrai}
\thanks{These two authors contributed equally}
\affiliation{Department of Physics, Bar-Ilan University, Ramat Gan, Israel}
\author{Nagendra K. Panduranga}
\affiliation{Department of Physics, Boston University, Massachusetts, USA}
\author{Reuven Cohen}
\affiliation{Department of Mathematics, Bar-Ilan University, Ramat Gan, Israel}
\author{Shlomo Havlin}
\affiliation{Department of Physics, Bar-Ilan University, Ramat Gan, Israel}
\affiliation{Institute of Innovative Research, Tokyo Institute of Technology, Midori-ku, Yokohama 226-8503, Japan}

\begin{abstract}
Modularity is a key organizing principle in real-world large-scale complex networks. Many real-world networks exhibit modular structures such as transportation infrastructures, communication networks and social media. Having the knowledge of the shortest paths length distribution (DSPL) between random pairs of nodes in such networks is important for understanding many processes, including diffusion or flow. Here, we provide analytical methods which are in good agreement with simulations on large scale networks with an extreme modular structure. By extreme modular, we mean that two modules or communities may be connected by maximum one link. As a result of the modular structure of the network, we obtain a distribution showing many peaks that represent the number of modules a typical shortest path is passing through. We present theory and results for the case where inter-links are weighted, as well as cases in which the inter-links are spread randomly across nodes in the community or limited to a specific set of nodes. 

\end{abstract}

\maketitle 

\section{Introduction}
The study of complex networks gains extensive interest in the last years as networks successfully model and lead to better understanding of many real world systems and processes in which interacting objects are involved.
In these models, objects are represented as nodes, and the interactions by links \cite{albert2002statistical, caldarelli2007scale, cohen2010complex, watts1998collective, newman2018networks, barabasi2016network, estrada2012structure}.

\begin{figure}[h!]
	\includegraphics[width=0.99\linewidth]{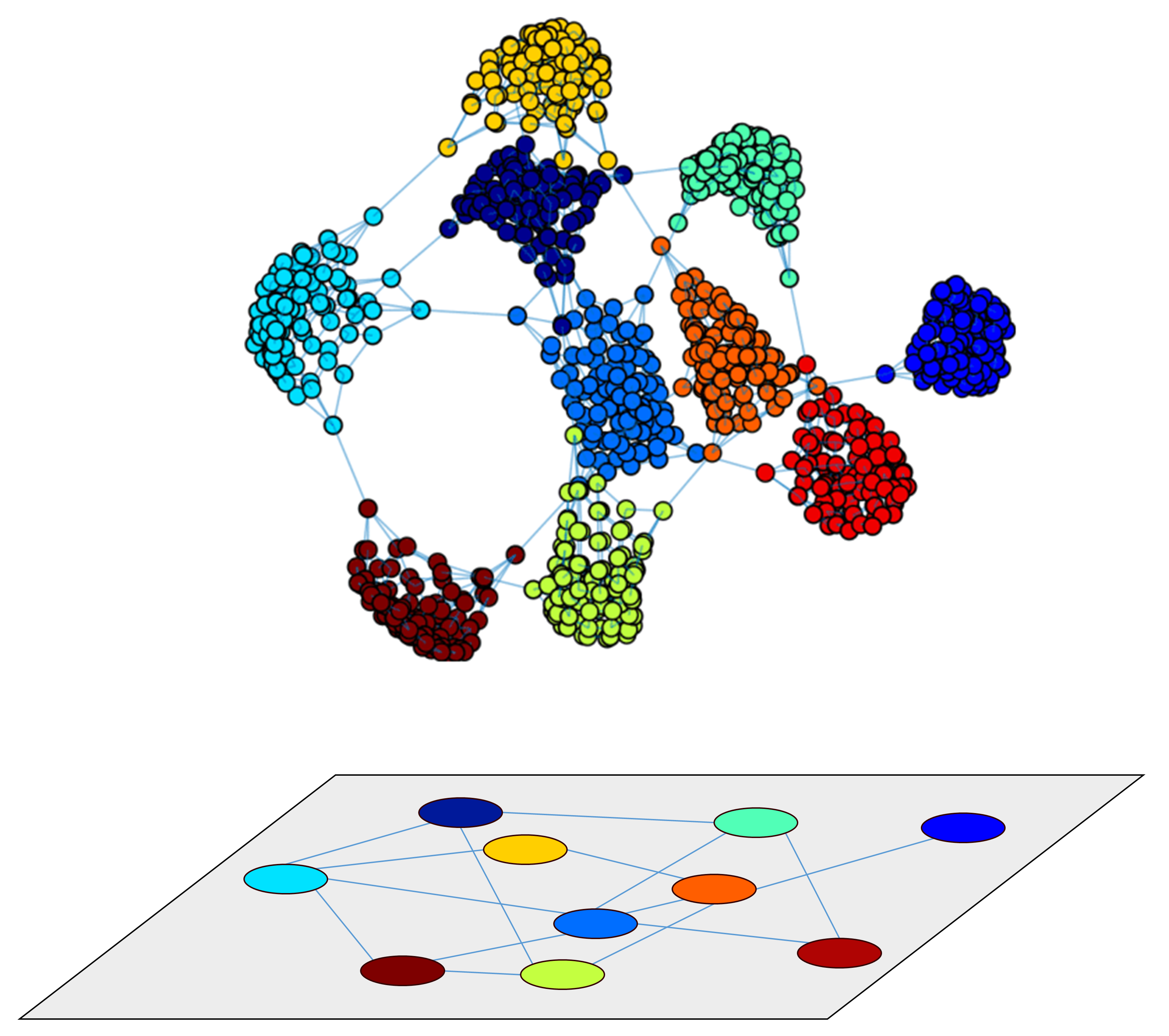}
	\caption{{\bf Illustration of the model.} We study networks that comprise $M$  modules, each of which has the same degree distribution, same topology and same size, $n$. Thus, we study a modular network with a total of $M \times n=N$ nodes.
			The upper picture shows the whole network, where nodes of each module have different color. The lower picture, is a projection of the upper network, showing the outer network, where each node represents a module. 
			 We start by constructing each module according to a given topology, mean degree ($k_{in}$) and links distribution, and  then construct the outer network which has its own topology, mean degree ($k_{out}$) and links distribution, where we consider each module as a node. Two modules can be connected by a maximum of one link. In section \ref{sec: results} we provide an analysis of different network topologies.
	}
	\label{fig: m comm illustration}

\end{figure}

Many real world networks exhibit a modular or community structure \cite{Eriksen_internet_2003,Guimer2005,garas2008structural,bullmore2012economy}. That is, a network is comprised of smaller networks (called communities, or modules) that are highly connected within themselves (by intra-links), and have a lower number of links between them (inter-links), which is a key to their structure and function. For demonstration, see Fig. \ref{fig: m comm illustration}. Knowing the distances distribution within networks with such topology is important for many reasons such as designing fast-communication, navigation, disease spreading and for optimizing processes on large graphs.

For each random pair of nodes $i$ and $j$ in the network, many paths can exist, or non at all. The distance between a pair of nodes is naturally defined as the shortest path length among all the paths existing between them. Distribution of shortest paths are expected to depend on the network structure and size. However, apart from a few studies \cite{NewmanWattsStog2001, DOROGOVTSEV2003,  Hofstad2005RandomStruct, vandenEsker2005Extremes, Hofstad_distPL_2008JMP, van2001paths}, the shortest paths length distribution (DSPL) despite its importance, attracted little attention. Recent studies developed analytical methods to compute the DSPL in \er \space  and configuration-model networks  \cite{katzav2015analytical, nitzan2016distance}. Another paper studied the DSPL in modular random networks \cite{dorogovtsev2008organization},  testing the conditions in which the number of inter-links between two or more modules control the network topology. This means,  answering the question "how many links between two modules are needed in order to unite them into one?". Adding more inter-links results in a change of the SPL distribution, which approaches a $\delta$ function as we add more inter-links.
Still, the case where the connections between the modules is itself a complex network, meaning that the inter-links are determined according to a given outer network, an analytical approach for finding the DSPL has not been developed yet.

\begin{figure}[h]
	\includegraphics[width=0.99\linewidth]{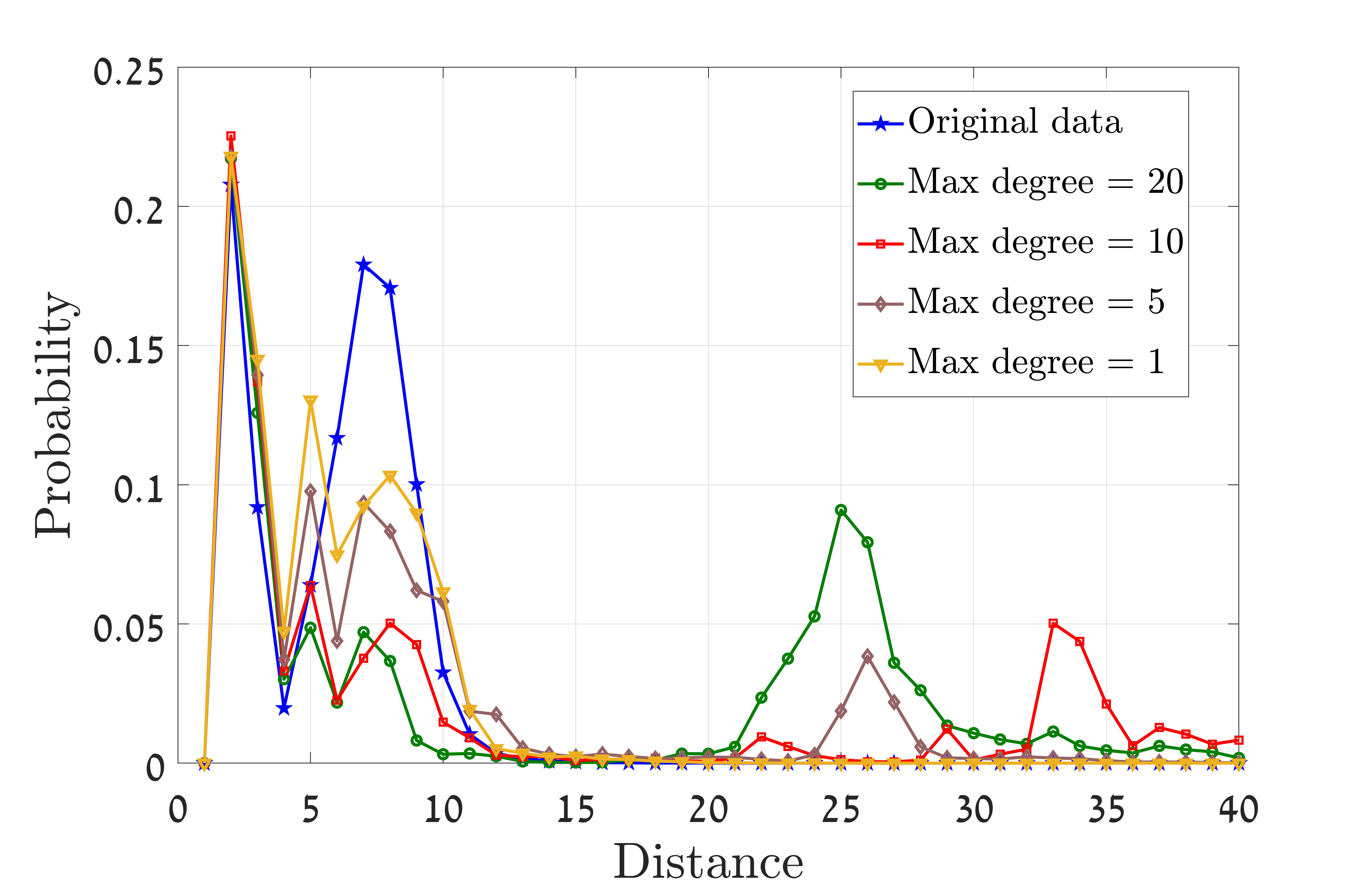}
	\caption{{\bf Analysis of autonomous systems (AS) distances distribution for different values of maximal number of inter-links outgoing from one module (max degree).} AS data was obtained from Caida and was collected using MIDAR-iff, where router topology based on aliases discovered by MIDAR, iffinder, and kapar. 
		 The data contained approximately 100 million nodes which are assigned into ~47,000 communities (AS). a node that connects two different ASs is an inter-linked node. Here we consider a case in which the Internet network undergoes a deliberate attack or experience node failures due to lack of electricity supply on inter-links between AS. We expect that inter-linked nodes fall with higher probability, due to their high betweenness centrality (in case of a deliberate attack) or high power demand in the case of power outage). We examine the distance distribution 
		 for those cases, allowing different maximal inter-link degree of an AS. Here, distance is defined as the shortest path length between two nodes in the graph. Note the wavy pattern of the distribution.}	
	\label{fig: AS_dist}
\end{figure}

As a motivation for the present study, we analyzed the distance distribution (DSPL) in the Internet routers-IP autonomous systems (AS) network. Each AS functions as a community, and contains routers IPs which are the nodes inside the community. The data were obtained from the center of applied Internet analysis (Caida) \cite{Caida}. Several studies have been performed on distances in the Internet \cite{albert1999internet, canbaz2017analysis, zhou2004accurately}, here we want to point out a specific phenomena which occurs when more and more inter-links are removed. In this case, a wavy distribution emerges. In Fig.~\ref{fig: AS_dist} we show the distance distribution (DSPL) of our data for different values of maximal inter-links degrees. By limiting the number of inter-links of an AS, we mimic a situation in which the Internet network undergoes an attack or power shortage. We can observe in Fig. \ref{fig: AS_dist} multiple peaks for the DSPL after such an attack, representing the modules passed by the shortest path. This phenomena motivates us here to develop a simplified model of extreme community structure which exhibits a wavy DSPL, and we study this analytically in order to better understand this phenomena.

In this paper we develop an analytical approach for obtaining the DSPL of a modular network. 
Our theory calculates the DSPL of the network given the shortest path distribution within one module (in net), and the DSPL of the network that connects the modules by treating each module as a node (out net). Our method holds for any inner and outer network topologies, and not only for random networks. The model we suggest assumes an extreme community condition where each module $I$ is connected to module $J$ with a maximum of one link that connects two randomly chosen nodes in both modules. Another condition we assume here is that the outer network has no small loops as explained in detail below.
In order to better simulate real world phenomena, such as routing and transportation between cities or countries, our model assumes a weight $w$ for inter-links, where intra-links weight is set to 1.
We further include analysis of various cases in which inter-links are limited to a specific set of nodes rather than being chosen randomly from the inner network. Analytic analysis of specific network topologies is also included.

The paper is organized as follows: In Sec. \ref{basicModel}  we present the basic model and theory we use to find the DSPL. In sections \ref{sec: one interconnected node} and \ref{sec: DifferentCasesOfConnectionsTours} we extend the theory for  
different inter-links configurations.
In Sec.~\ref{sec: results} we present our results of DSPL, from both our theory and simulations of selected network topologies. More comprehensive mathematical analysis of some specific cases of network of networks is presented in more detail in the Appendix.

\section{Model and Theory}
\label{modelTheory}

\subsection{Basic model}
\label{basicModel}
Let a network consist of $m$ communities, or modules. Each module is assumed to be of the same size and constructed in the same fashion (or just with the same distances distribution), e.g., \er, scale-free (SF), random regular (RR), lattice or any other structure. An outer network, which also can take any structure, regards every module as a node.  Therefore we obtain a "large" network which comprises modules, and another network on top of it which connects those modules as illustrated in Fig.~\ref{fig: m comm illustration}. \\

\noindent
Our model assumes the following: \\
$\mathit{1}$. There is at most one inter-link between two modules. \\
$\mathit{2}$. The inter-links connect between pairs of \textit{random} nodes of two modules. \\
$\mathit{3}$. inter-links have a weight $w$ (integer), while the weight of intra-links is 1.
\\
$\mathit{4}$. The outer network has no small loops. 
\\
$\mathit{5}$. As a consequence of $\mathit{4}$, an outer shortest path between modules in the outer network is single and the second outer shortest path is much longer than it. Therefore, the shortest path in the whole network, in most cases, will pass through the shortest path of the outer network. \\

\noindent
It is important to notice that while assumptions 4 and 5 hold for short distances, they partially fail for the long distances in the network. Hence, we expect slight deviations at the end of the distribution, as seen in general in the figures. Random sparse networks, for instance, exhibit locally tree-like behavior \cite{newman2018networks382}. The range of this behavior is up to the average distance of the network approximately \cite{cohen2010complex12}, therefore for these networks our theory is accurate up to the average distance of the network, and then it has slight deviations as we show below. For 1D lattice, for example, assumptions 4 and 5 are valid up to the longest distances, whereas for a 2D lattice, the assumptions fail.  
\\

\noindent
%Now, let a network which consists of modules, such that 
Now, the shortest path length (SPL) distribution in each module (inner paths) is $P_l^{in}$,  and has the generating function $G_{in}(x) = \sum_{l=0}^{\infty}P_l^{in}x^l $. Likewise, the SPL distribution of the outer network is $P_l^{out}$ and has the generating function $G_{out}(x)= \sum_{l=0}^{\infty}P_l^{out}x^l$. \\

\begin{figure}[h]
	\includegraphics[width=0.99\linewidth]{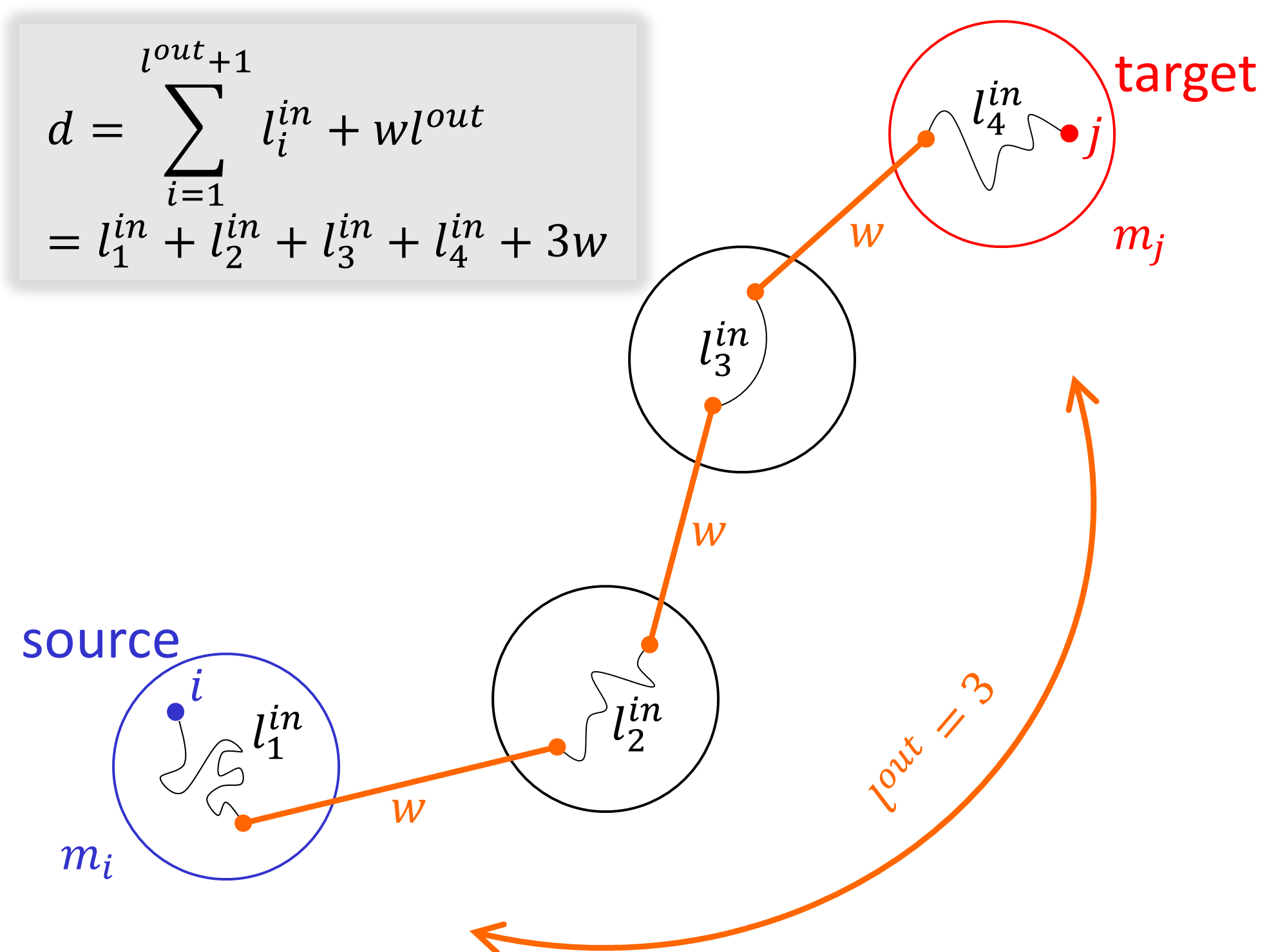}
	\caption{{\bf Illustration of the problem and the theory.} Consider two random nodes $i$ and $j$, which reside in different modules $m_{i} $ and $m_{j}$. In order to reach via the shortest path from node $i$ (source) to node $j$ (target), one has to walk as follows. First, to find the outer shortest path that connects the modules ($l^{out}$). Next, to look for the shortest path within $m_i$ to the node that connects the source node to a node that resides in the next module of $l^{out}$, which is denoted by $l_1^{in}$ in the figure. We iterate this process again in the next modules on the path, until we finally land in our target node. Our total path length will be $d = \sum{l_i^{in} + wl^{out} }$.
	 }
		
	\label{fig: Theory illustration}
\end{figure}

\noindent
According to the above assumptions, one can find that the SPL between two random nodes in the network satisfies
\begin{equation}
	d = \sum_{i=1}^{l^{out}+1}l_i^{in}+wl^{out},
	\label{eq: d first}
\end{equation}
which yields
\begin{equation}
d + w = \sum_{i=1}^{l^{out}+1}(l_i^{in}+w) ,
\end{equation}
where $d$ is the total distance between two random nodes, $l^{out}$ is the external shortest path length between the communities those nodes reside in, and $l_i^{in}$ is the internal distance between nodes in the same community which function as the connecting nodes between the communities in $l^{out}$. See illustration in Fig.~\ref{fig: Theory illustration}. \\

\noindent
This is a sum of independent random variables where the number of elements of the sum itself, is also a random variable.
Then we can use known theorems \cite{johnson2005univariate} to conclude the following results. \\
First, from Wald’s identity,
\begin{equation*}
\av{ d + w } = \av{l^{out}+1}\av{l^{in}+w},
\end{equation*}
which gives
\begin{equation}
\av{ d } = (\av{l^{out}}+1)(\av{l^{in}}+w) - w.
\label{eq: average distance}
\end{equation}
This result suggests that in small world networks the extreme modularity condition makes the average distance much longer.
Furthermore, for the generating functions one can write
\begin{gather*}
\big[x^wG_d(x)\big] = \big[xG_{out}(x)\big]\circ \big[x^wG_{in}(x)\big] \\
x^wG_d(x) = x^wG_{in}(x)G_{out}(x^wG_{in}(x)),
\end{gather*}
where $G_d(x)$ is the generating function of $P_d$, the probability distribution of $d$, and $\circ$ is a composition of functions.
\\
Thus, we get 
\begin{equation}
G_d(x) = G_{in}(x)G_{out}(x^wG_{in}(x)).
\label{eq: composition}
\end{equation}
Since we have the generating function of the shortest path distribution we are consequently able to find $P_d$ by derivation or integration (Cauchy formula) numerically by 
%\begin{gather*}
%P_d = \frac{G_d^{(l)}(0)}{d!} 
%\end{gather*}
%or by
\begin{gather}
P_d = \frac{G_d^{(d)}(0)}{d!} = \frac{1}{2\pi i}\oint \frac{G_d(z)}{z^{d+1}} \mathrm{d}z
\label{eq: integral}
\end{gather}
where the integral is performed on a close path around $z=0$ in the complex plain. This integral is far more simple to compute numerically than computing high derivatives. A simple contour can be a canonical circle with $r=1$.\\

\noindent
See Appendix \ref{sec: appendix specific} where we analyze analytically few specific cases of network of networks topologies. Including, 1-2D lattices, Poisson distance distribution, two modules and star graph. We find for these cases explicitly all or part of the following expressions. $G_{in/out}(x)$, $G_d(x)$ and $P_d$. For two modules with Poisson DSPL we find analytically also a condition for the appearance of two peaks rather than one peak, see Eqs.~\eqref{eq: ratio} and \eqref{eq: 2eqs criticality} and Figs.~\ref{fig: ratio} and \ref{fig: two poisson}.

%Finally, we mention that with respect to the outer network 
%\begin{gather*}
%k_{out}=nrk_{inter}
%\end{gather*}
%where $k_{out}$ is the mean number of neighbor communities of one community, $n$ is the number of nodes in each community, $r$ is the fraction of nodes can be connected to other community, $k_{inter}$ is the nean number of inter-links among the nodes can have such a link. \\

\subsection{One node has all the inter-links in each module} \label{sec: one interconnected node}
\begin{figure}[h]
	\includegraphics[width=0.4\linewidth]{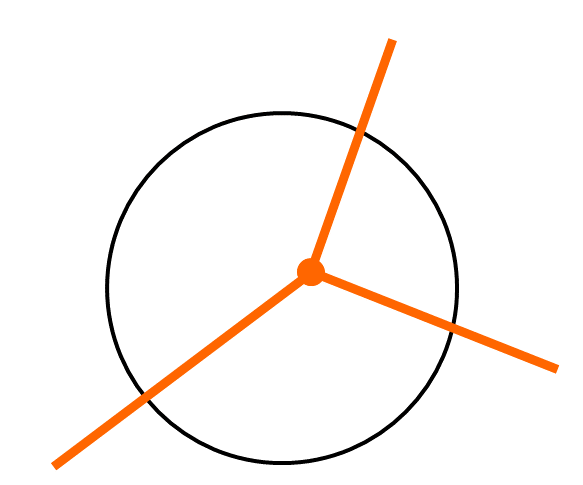}
	\caption{  {\bf Illustration of the model in section \ref{sec: one interconnected node}.} Here, each module has only one interconnected node to which all the inter-links of this module are connected.}
	\label{fig: illustration p=0}
\end{figure}

When analyzing the internet data (AS) that was mentioned above, we noticed the fact that many inter-connected nodes have multiple inter-links. In order to cover other realistic cases such as this, we consider also the scenario in which all the inter-links of a module go out and in from the same single inter-connected node, rather than from random nodes as the above model, see Fig.~\ref{fig: illustration p=0}. This situation changes the distance significantly,
\begin{equation*}
d = l_1^{in} + l_2^{in} + wl^{out},
\end{equation*}
and if $l^{out}=0$ then $d = l^{in}$ (because the source and the target reside in the same module).
Hence
\begin{equation}
\begin{aligned}
G_d(x) & = G_{out}(0)G_{in}(x) +
\\
&\left[ G_{out}(x^w)-G_{out}(0) \right] \left[ G_{in}(x) \right]^2.
\end{aligned}
\label{eq: same inter}
\end{equation}

\subsection{Different cases of inter-links connections}\label{sec: DifferentCasesOfConnectionsTours}
\begin{figure}[H]
	\centering
	\includegraphics[width=0.7\linewidth]{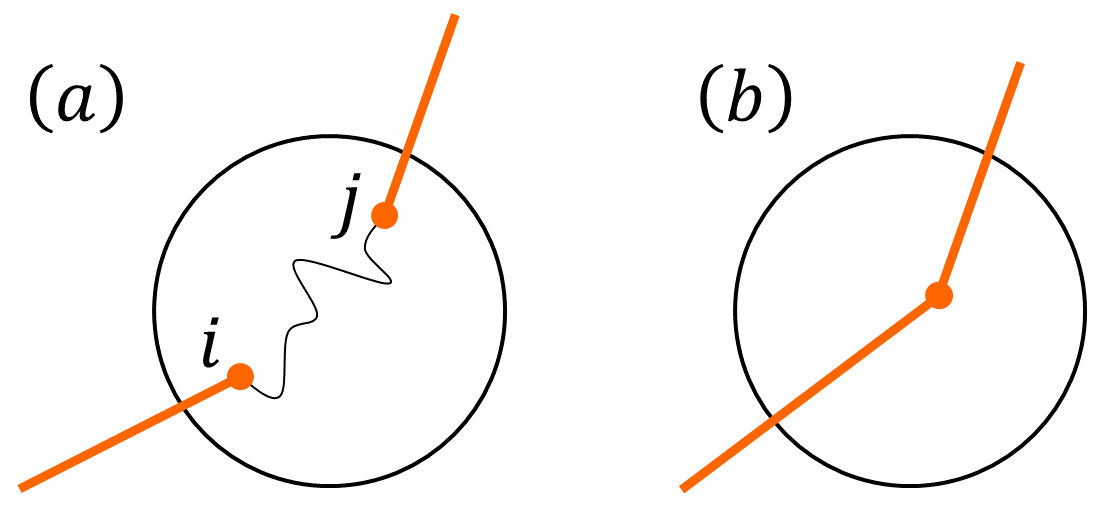}
	\caption{\textbf{Illustration of the model in section \ref{sec: DifferentCasesOfConnectionsTours}.} 
	 (a) In this scenario we enter the community through node $i$ and leave the community with probability $p$ through a different node $j$, with addition of a tour inside the community. (b) Here, with probability $1-p$, the arrival and the departure to and from the community is from the same node. }
	\label{fig: illustration lcon}
\end{figure}

In this section, we consider the case in which, when entering a module via an interconnected node $i$, we leave this module via different interconnected node $j$ with probability $p$, or, when departing the module via the same node with probability $1-p$.  See Fig. \ref{fig: illustration lcon}.

\begin{figure*}[ht]
	\includegraphics[width=0.99\textwidth]{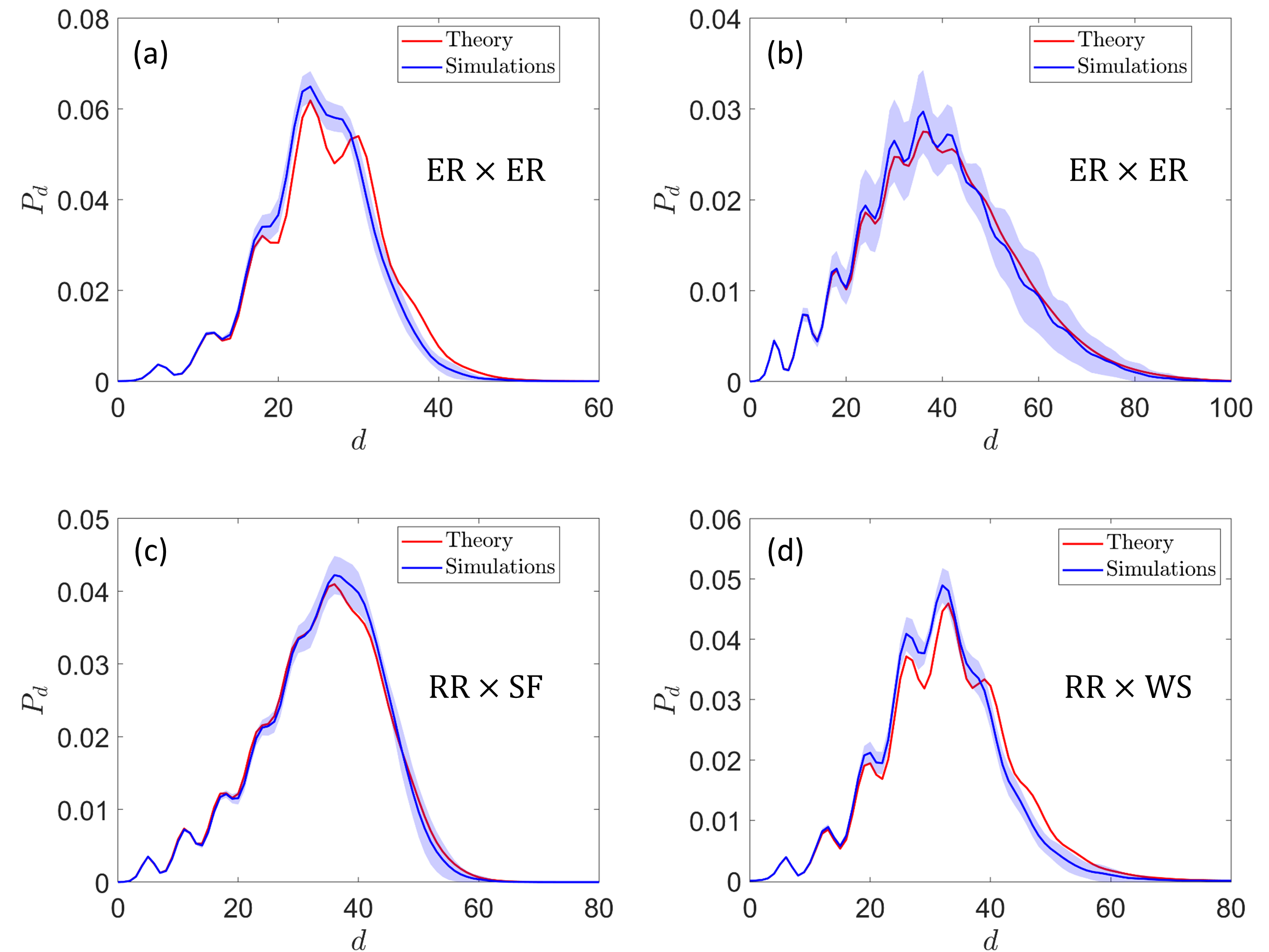}
	\caption{ 
		{\bf Results of distance distribution in several types of modular networks with uniform links weight ($w=1$).} In \textbf{(a)} we show the case of ER$\times$ER (outer network ER, modules ER), with $M=10^2$ (number of modules), $k_{out}=4,~n=10^3$ (size of each module) and $k_{in}=4$. Red line represents theory result, Blue line is the average of simulations, and shaded blue is the simulations standard deviation.
		\textbf{(b)} The same as \textbf{(a)} except that $k_{out}=2$. In \textbf{(c)} we show RR of SF (out: RR, in: SF) where $M=10^2,~k_{out}=3,~n=10^3,~k_0^{in}=2$ and the power law degree exponent is $\gamma_{in}=3$. In \textbf{(d)} we show RR of WS (Watts–Strogatz model) where $M=10^2,~k_{out}=3,~n=10^3,~k_{in}=4$ and $\beta=0.5$. Simulation results were taken over 10 realizations. In all cases one can see a good agreement between theory and simulations except for slight deviations at large distances, the reason of which is discussed in the text. The wavy distributions found here are significantly wider than a distribution of a single network, due to the extreme modular structure. In a single ER network, the mean distance is given by $\av{d} \approx {\ln(N)}/{\ln(k)}$, where in our model in the case of ER$\times$ER the mean distance is $\av{d} \approx (\ln(n) / \ln(k_{in})+1) (\ln(M)/\ln(k_{out})+1)-1$. See Eq.~\eqref{eq: average distance}.
	}
	\label{fig: m comm}
\end{figure*}

\noindent
In appendix \ref{sec: DifferentCasesOfConnectionsTours appendix} we find that for this case
\begin{equation}
\begin{aligned}
G_d(x) & = G_{out}(0)G_{in}(x) +
\\
&  \left[ G_{in}(x) \right]^2 \frac{  G_{out}\left( x^w \big( 1-p + pG_{in}(x) \big) \right)-G_{out}(0) }{ 1-p + pG_{in}(x) }.
\end{aligned}
\label{eq: p-model}
\end{equation}
One can see that the last equation converges nicely to those of chapters \ref{basicModel} (Eq.~\eqref{eq: composition}) and \ref{sec: one interconnected node} (Eq.~\eqref{eq: same inter}) at the limits $p=1$ and $p=0$ respectively.

\section{Results}
\label{sec: results}

\begin{figure*}[ht]	
	\includegraphics[ width=0.99\textwidth]{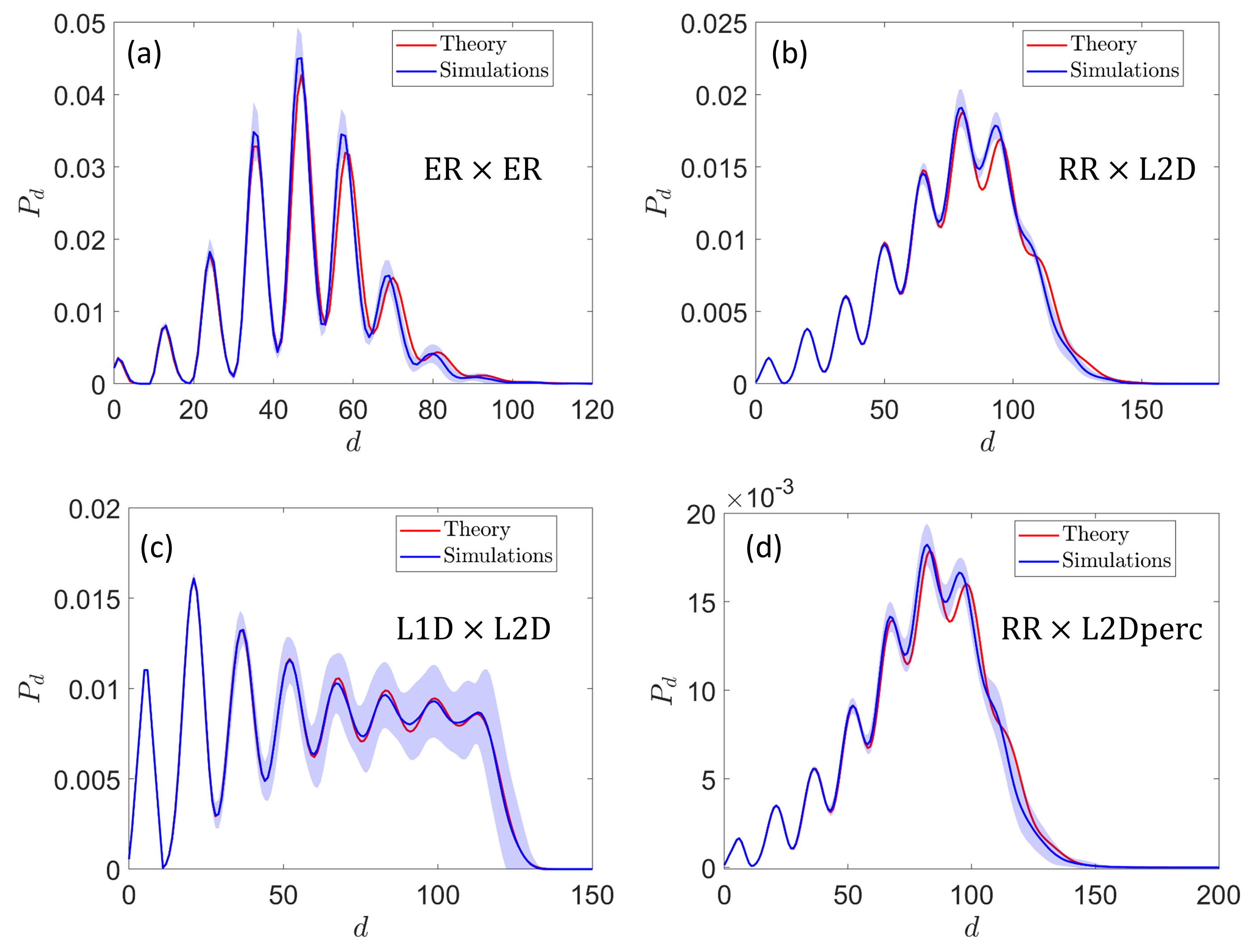}
	\caption{
		{\bf Weighted inter-links cases.} Assuming a weight $w$ to inter-links, results for random and lattice networks are shown. In  \textbf{(a)} we show a ER of ER network with the parameters $M=10^2,~k_{out}=3,~w=10,~n=10^3$ and $k_in=3$. In  \textbf{(b)} we show a RR of L2D (outer: RR, inner: 2D lattice) network with the parameters $M=10^2,~k_{out}=3,~w=10$ and $n=10^2$. In \textbf{(c)} we show L1D of L2D (outer: 1D lattice, inner: 2D lattice) with $M=15,~w=10$ and $n=121$ and the theory is from the explicit formula in the appendix. In  \textbf{(d)} we show a RR of L2Dperc (2D lattice with percolation where fraction $q$ of random nodes was removed) network with the parameters $M=10^2,~k_{out}=3,~w=10,~n=10^2$ and $q=0.2$. For this case, we test the DSPL in the giant connected component. 	
		All cases show exclusive distances distribution and good agreement between theory and simulations.
	}
	\label{fig: m comm with w}
\end{figure*}

Fig.~\ref{fig: m comm} compares between theory and simulations for different network layouts and parameters where the inter-links have the same length as the inner links, i.e. $w=1$.
In general the  figure shows a good agreement between theory and simulations. \\
It is important to notice the distance distribution exhibits a wavy behavior on top of a hill envelope. The intuitive explanation for this is that each hill represents paths between nodes in two modules that have the same outer distance. The first hill comes from paths between nodes inside the same module, while the second hill comes from paths between neighboring modules, which are about twice longer due to their consistency of two inner paths - the first, in the source module, from the source node to the interconnected node inside the source module,  and the second, from the interconnected node in the target module, to the target node. The second hill is higher because there are more paths between neighboring modules than paths within a single module. In other words, in the outer network (in between modules), there are more shortest paths with $l^{out}=1$ than with $l^{out}=0$. The same holds for the third hill ($l^{out}=2$) and so on. That is to say, what rules the hills' heights is the outer SPl distribution, therefore we get a bell shaped envelope which comes from the outer network distribution, and upon it hills which come from inner networks distribution.
%\red{ Another perspective, is that the wavy pattern is an outcome of a correlations of many paths, meaning, the fact that many paths share an overlapping sub paths. For example, all the paths from module $m_i$ to module $m_j$, share a large common route. All the shortest paths from nodes that reside in $m_i$ to nodes that reside in $m_j$, share a common path on the way from $m_i$ to $m_j$ (See Fig. \ref{fig: Theory illustration}). Those correlations bundle the paths to a separate sets which are expressed in the emergence of the wavy distribution.}
\\

\noindent
Note that, for the long distances there is a slight deviation between the theory and the simulations results. 
This can be explained by the fact that in theory we neglect loops in the outer network, while in practice, for finite networks, there are long loops (the short ones are negligible). The long loops causes that there are modules far from each other have few outer similar paths between them.
This multiplicity of similar outer paths shortens the distance from a source node to a target node because the shortest path is chosen among them. In this case, we will need to find the minimum of similar independent random variables, which is different (lower) than the expectation value relative to the random variables. 

Fig.~\ref{fig: m comm with w} shows the results for the distance distribution for the extreme modular network, both theory and simulations, where inter-links are weighted with $w=10$. It can be seen that the separation between the hills becomes more significant because paths between modules with different outer distances have dramatically different lengths as a result of the length of the inter-links. Within a single 2D lattice there is a broad distance distribution because the system is not a small world network. As a result when the inter-links are not much longer than the inner ones, the wavy behavior vanishes because the widths of hills are large so they become blended together. However, when $w$ is sufficiently large, the waves are very distinct.

\begin{figure*}[ht]	
	\includegraphics[ width=0.99\textwidth]{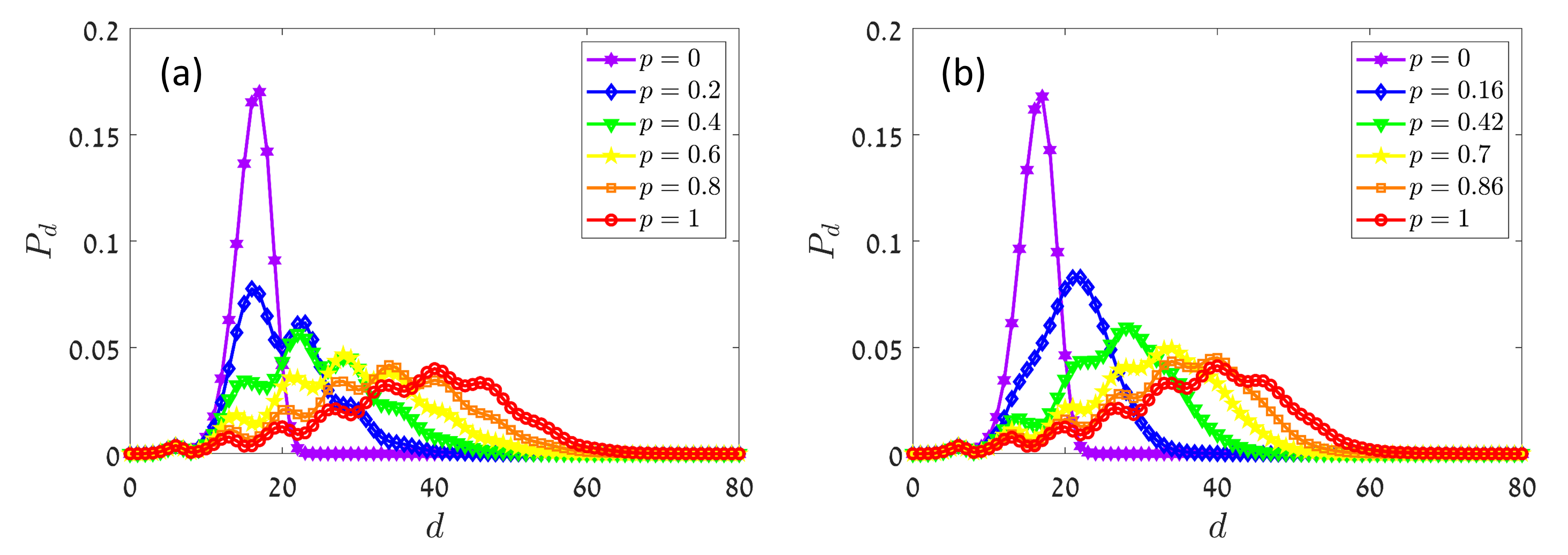}
	\caption{
		{\bf Impact of the parameter $p$ in both theory (a) and simulations (b).} Results of Eq.~\eqref{eq: p-model} where the network is formed as follows. Out: RR, $M=100,~k=3,~w=1$. In: RR, $n=1000,~k=4$. where (a) shows the theory, Eq.\eqref{eq: p-model}, and (b) shows simulation results. The values of $p$ in simulations were found by sampling many realizations of random shortest paths and measuring how many times each path goes in and out a module through different interconnected nodes and how many times via the same node. Results show good agreement between theory and simulations. 
	}
	\label{fig: r results}
\end{figure*}

Fig. \ref{fig: r results} shows the results of Eq.\eqref{eq: p-model}, for various values of $p$. This model  suits a more realistic case, in which there is a  probability $p$ of accessing and leaving a community through a different or the same ($1-p$) interconnected node. Note the reduction in the number of waves when $p$ approaches 0, which is the case in which no intra-module paths were taken.
%\subsection{Small outer and inner degrees close to criticality}
%In ER network it is known that $\langle l \rangle \sim \log N$, but where $k\to k_c$ it is satisfied that $\langle l \rangle \sim N^{1/3} ?$. \\
%Hence the average distance satisfies
%\begin{gather*}
% \av{d} \sim \log m \log n
%\\
% \text{or}
% \\ 
%\av{d} \sim  m^{\frac{1}{3}} \log n
%\end{gather*}
%In Fig. \ref{fig: ER_exponent} we show $\langle d \rangle$ vs $m$ (number of communities) for different values of degree $k$.

%\begin{figure}[ht]
%	\includegraphics[height=7cm, width=0.49\textwidth]{exponent_ER_ER.png} 
%
%	\caption{
%		{\bf exponent for ER networks.} 
%	\red{Do we keep* it or no?}
%	}
%	\label{fig: ER_exponent}
%\end{figure}

\begin{figure*}[ht]	
	\centering
	\includegraphics[width=0.49\linewidth]{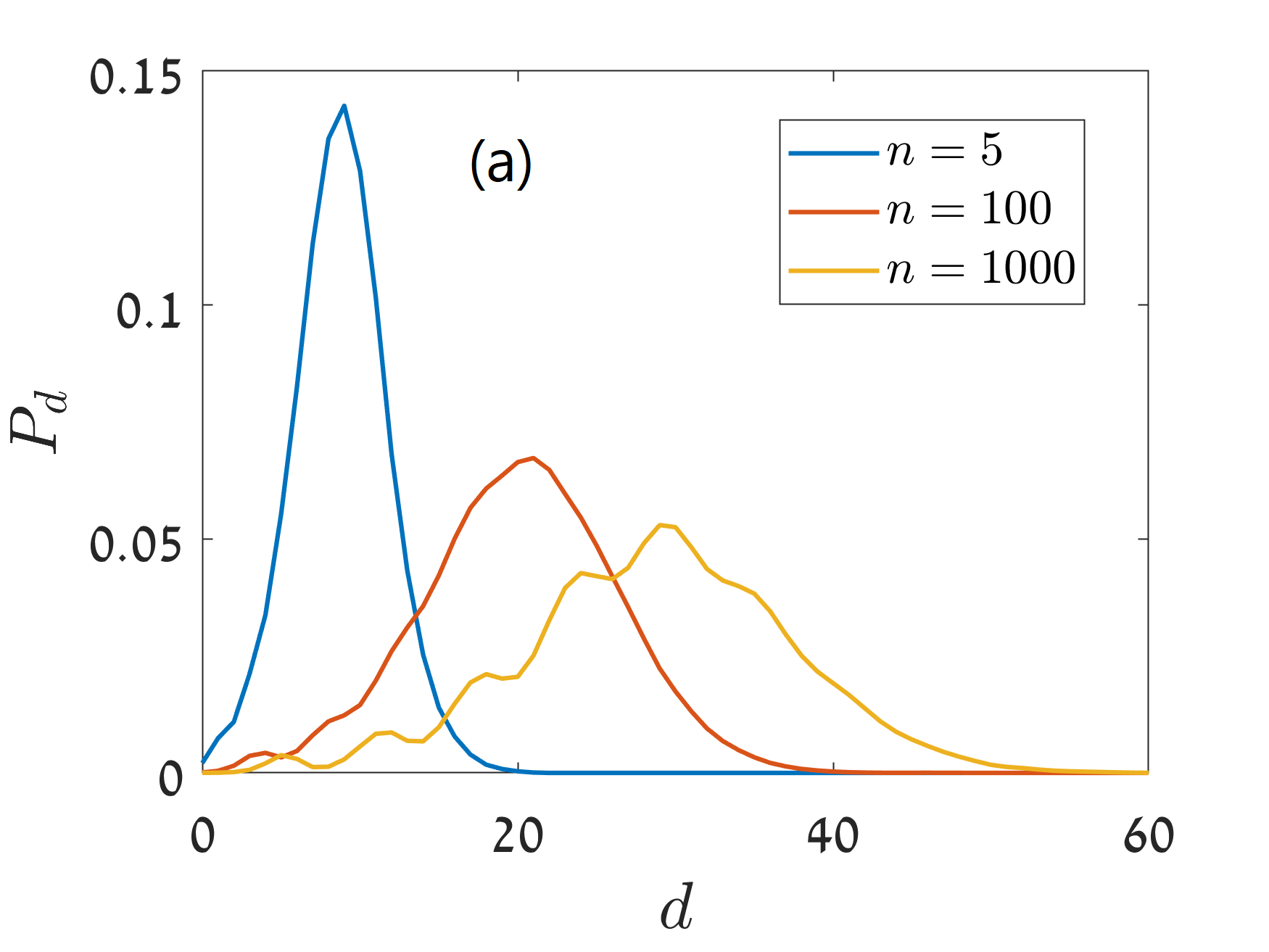}
	\includegraphics[width=0.49\linewidth]{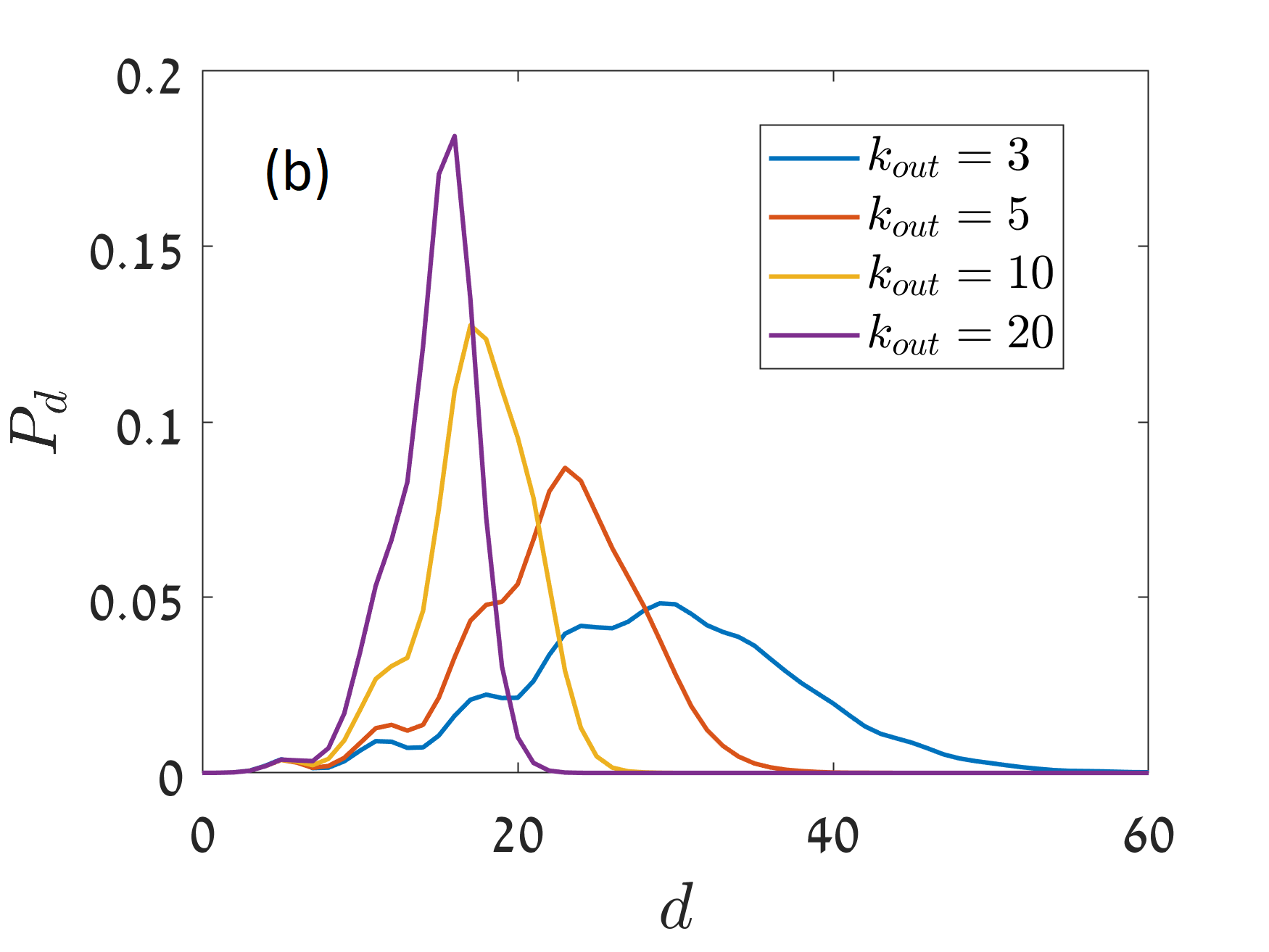}	
	\caption{
		{\bf Impact of the modules size and the outer degree on the wavy distribution.} In (a) the network parameters are Out: ER, $M=100,~k=3,~w=1$. In: ER, $k=4$ and $n$ changes. The results were averaged over 5 realizations of simulation. In (b) all the parameters are the same, except $n=1000$ and $k_{out}$ changes.
	}
	\label{fig: waves emergence}
\end{figure*}

In order to examine the emergence of the wavy distribution, we regulate the parameter $n$, modules size. We show in Fig.~\ref{fig: waves emergence}a that where $n$ is very small the network acts as a single network, of course. However, when we increase $n$ more and more, at some point the wavy pattern appears and becomes more and more clear. In the Appendix Sec. \ref{sec: two modules app}, we find analytically a criterion for the emergence of a wavy distribution in a network of two modules. In Fig.~\ref{fig: waves emergence}b we show by changing the outer average degree, $k_{out}$, how the sparsity of the outer network affects the waviness of the distribution.

\section{discussion}
In this paper we develop a framework to find  analytically the distance distribution within networks with extreme community structure given the distributions of the inner and outer networks. We study here a model where we assume there is at most a single inter-link between modules. 
We showed that the SPL distribution has a wavy pattern in good agreement with simulations. Future work can investigate the validation of this model for real networks, where multiple links between modules exist.

\section{Acknowledgments}

We thank the Italian Ministry of Foreign Affairs and International Cooperation jointly with the Israeli Ministry of Science, Technology, and Space (MOST); the Israel Science Foundation, ONR, the Japan Science Foundation with MOST, BSF-NSF, ARO, the BIU Center for Research in Applied Cryptography and Cyber Security, and DTRA (Grant no. HDTRA-1-10-1-0014) for financial support.

\bibliographystyle{unsrtnat}
\bibliography{communities}

\clearpage

\setcounter{figure}{0}
\setcounter{equation}{0}
\renewcommand{\thefigure}{A.\arabic{figure}}
\renewcommand{\theequation}{A.\arabic{equation}}

\appendix
\section{Specific networks} \label{sec: appendix specific}

\subsection{1D lattice}
Consider a 1D lattice with periodic boundaries with size $L$. For simplicity we take odd $L$. Then, the distances of each node from all other nodes have the following frequency
\begin{gather}
N_l = 
\begin{cases}
1, \hspace{0.7cm} l=0
\\
2, \hspace{0.7cm} 1 \leq l \leq (L-1)/2
\\
0,  \hspace{0.72cm} l>L/2
\end{cases},
\label{eq: Nl 1D lattice}
\end{gather}
where $N_l$ is the number of nodes in distance $l$ from the source node. Then, the distance distribution, $P_l$, is obtained by
\begin{equation}
P_l = N_l/L.
\end{equation}
The generating functions of $N_l$ and $P_l$ satisfy
\begin{equation}
\begin{aligned}
&
\begin{aligned}
G_{N}(x) &= -1+2\left(1+x+...+x^{(L-1)/2}\right) 
\\
&= -1+2 \frac{1-x^{(L+1)/2}}{1-x} ,
\end{aligned} \\[8pt]
&G_P(x)=G_N(x)/L.
\end{aligned}
\end{equation}
Thus,
\begin{equation}
G_{1D}(x) = G_P(x) = \frac{1}{L} \left(-1+2 \frac{1-x^{(L+1)/2}}{1-x}\right).
\label{eq: G(x) 1d lattice}
\end{equation}

\noindent
Comment: In Eq.~\eqref{eq: Nl 1D lattice} we counted twice the distances between different nodes $i$ and $j$ ($(i,j)$ and $(j,i)$), and only once the distance (0) between a node $i$ to itself. The reason is that we define $P_l$ as the probability of the distance between two random nodes to be $l$. Indeed the probability to choose different nodes $i$ and $j$ is twice as large as the probability to choose the same node $i$ twice. Note, it matters only for the value of $P_0$.

\subsection{2D lattice}
Consider a 2D square lattice with periodic boundaries and size $L \times L$. We assume for simplicity that $L$ is odd. Then, the distances of each node from all other nodes have the following frequency
\begin{equation}
N_l = 
\begin{cases}
1, & l=0
\\
4l, & 1 \leq l < L/2
\\
4(L-l), & L/2 < l \leq L 
\\
0,  & l>L
\end{cases},
\end{equation}
and the distance distribution is
\begin{equation}
P_l = N_l/L^2.
\end{equation}
We note that $N_l$ is obtained by a convolution of the series $a_l$ and $b_l$, where
\begin{equation}
a_l= 
\begin{cases}
1, & 0 \leq l < L/2-1
\\
0, &l > L/2-1
\end{cases} , \quad
b_l= 
\begin{cases}
1, & 0 \leq l < L/2
\\
0, & l > L/2
\end{cases} ,
\end{equation}
such that
\begin{equation}
\begin{cases}
N_0 = 1
\\
N_{l+1} = 4(a_l * b_l)
\end{cases}.
\end{equation}
As a result, the generating functions of these sequences ($a_l,~b_l,~N_l,~P_l$) satisfy 
\begin{equation}
\begin{cases}
G_{N}(x)=1+4xG_{a}(x)G_b(x)
\\
G_P(x)=G_N(x)/L^2
\end{cases}.
\end{equation}
But we note that
\begin{eqnarray}
\begin{aligned}
& G_{a}(x)= 1+x+...+x^{(L-3)/2} = \frac{1-x^{(L-1)/2}}{1-x}. 
\\
& G_{b}(x)= 1+x+...+x^{(L-1)/2} = \frac{1-x^{(L+1)/2}}{1-x}. 
\end{aligned}
\end{eqnarray}
Therefore, we obtain
\begin{equation}
\begin{aligned}
& G_{2D}(x) = G_P(x) 
\\
& = \frac{1}{L^2} \left(1+4x \frac{(1-x^{(L-1)/2})(1-x^{(L+1)/2})}{(1-x)^2}\right).
\end{aligned}
\label{eq: G(x) 2d lattice}
\end{equation}

\subsection{1D lattice of 2D lattices}
Consider a circle of square lattices that are interconnected with one inter-link between two random nodes from neighboring lattices. The inter-links have weight $w$ while the intra-links have weight 1. Then, the distance distribution is obtained according to Eq.~\eqref{eq: G(x) 1d lattice}, \eqref{eq: G(x) 2d lattice} and \eqref{eq: composition} by 
\begin{equation}
G_{1D\times 2D}(x) = G_{2D}(x)G_{1D}(x^w G_{2D}(x)) .
\end{equation}
This result is shown in Fig.~\ref{fig: m comm with w}c.

\subsection{Poisson distance distribution}
\begin{figure}[ht]	
	\centering
	\includegraphics[width=0.95\linewidth]{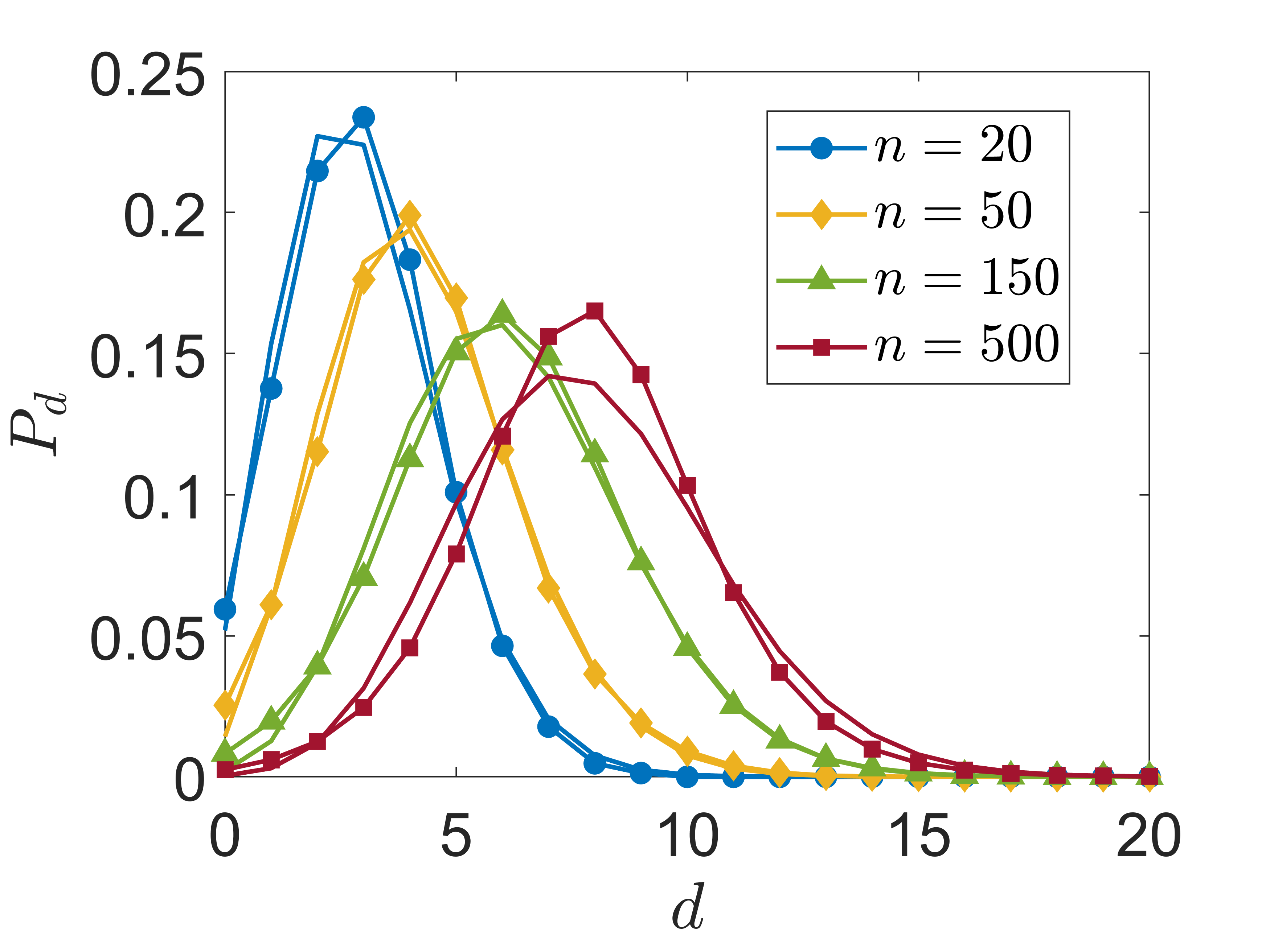}
	\includegraphics[width=0.95\linewidth]{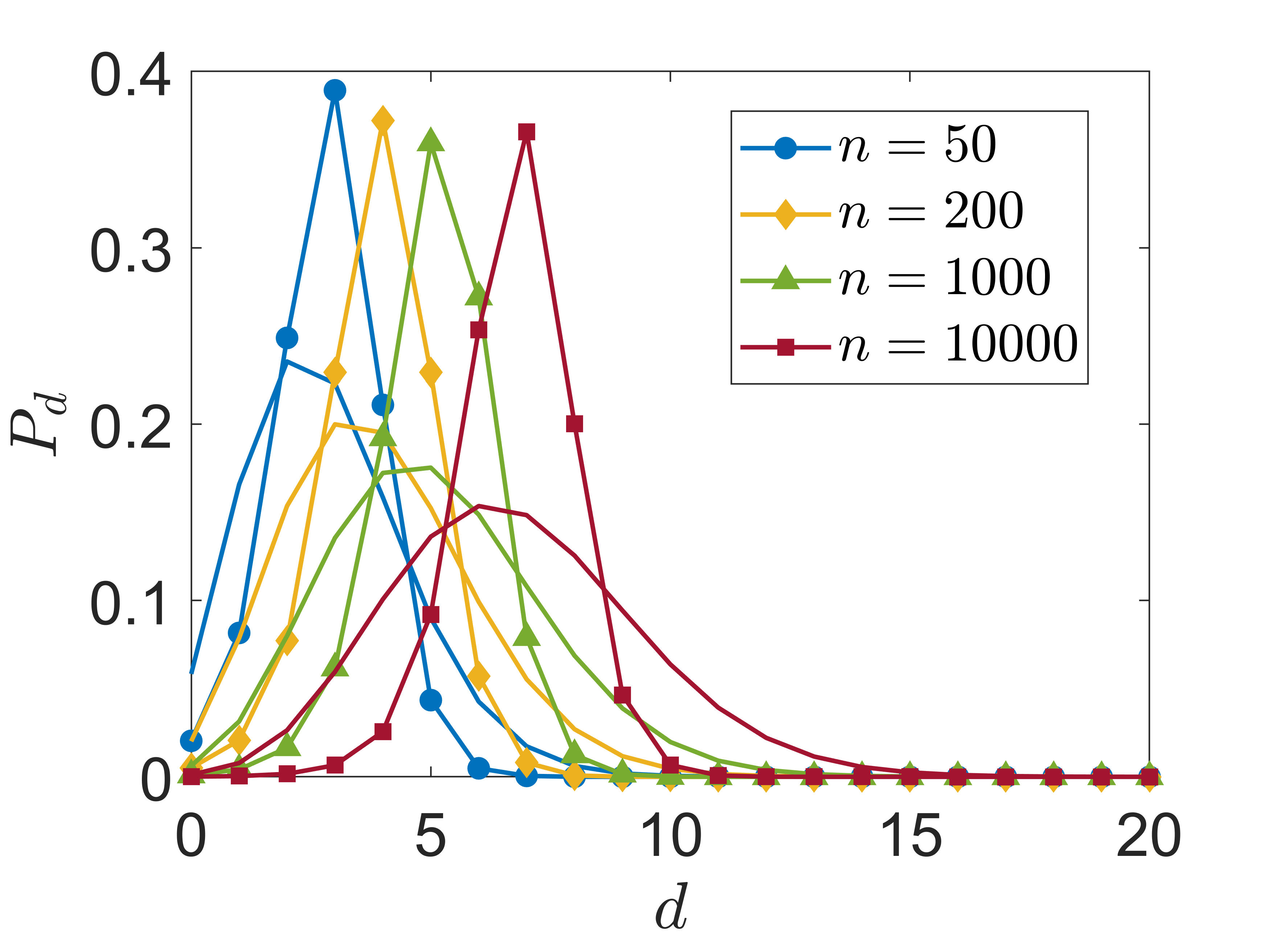}
	\caption{
		{\bf Test of Poisson approximation for ER DSPL.} The simulations are on ER network with $k=2$ (top) and $k=4$ (bottom). The lines with markers are from simulations and the lines without markers are Poisson distribution with the same average. One can see that for $k=2$ in this region of sizes the approximation is good, while if $n$ increases further it becomes less accurate. For $k=4$ it is much less accurate. Generally, we see a difference that Poisson distribution has a standard deviation $\sqrt{\lambda}$ where the average is $\lambda$, while the DSPL does not change its standard deviation while changing its average for large $n$.
	}
	\label{fig: ER vs poisson}
\end{figure}

To obtain insight into the wavy distribution, we assume here that the distances in a network have a Poissonian distribution. This will enable us to obtain analytically the distance distribution in our model of extreme modular networks. Indeed random networks have in certain parameters range a distance distribution which can be approximated by a Poissonian distribution, as shown in Fig.~\ref{fig: ER vs poisson}. This changes with the degree and the network size very much. For higher degrees it does not work so well, while for small degrees it works well. 

\begin{figure}[ht]	
	\centering
	\includegraphics[width=0.95\linewidth]{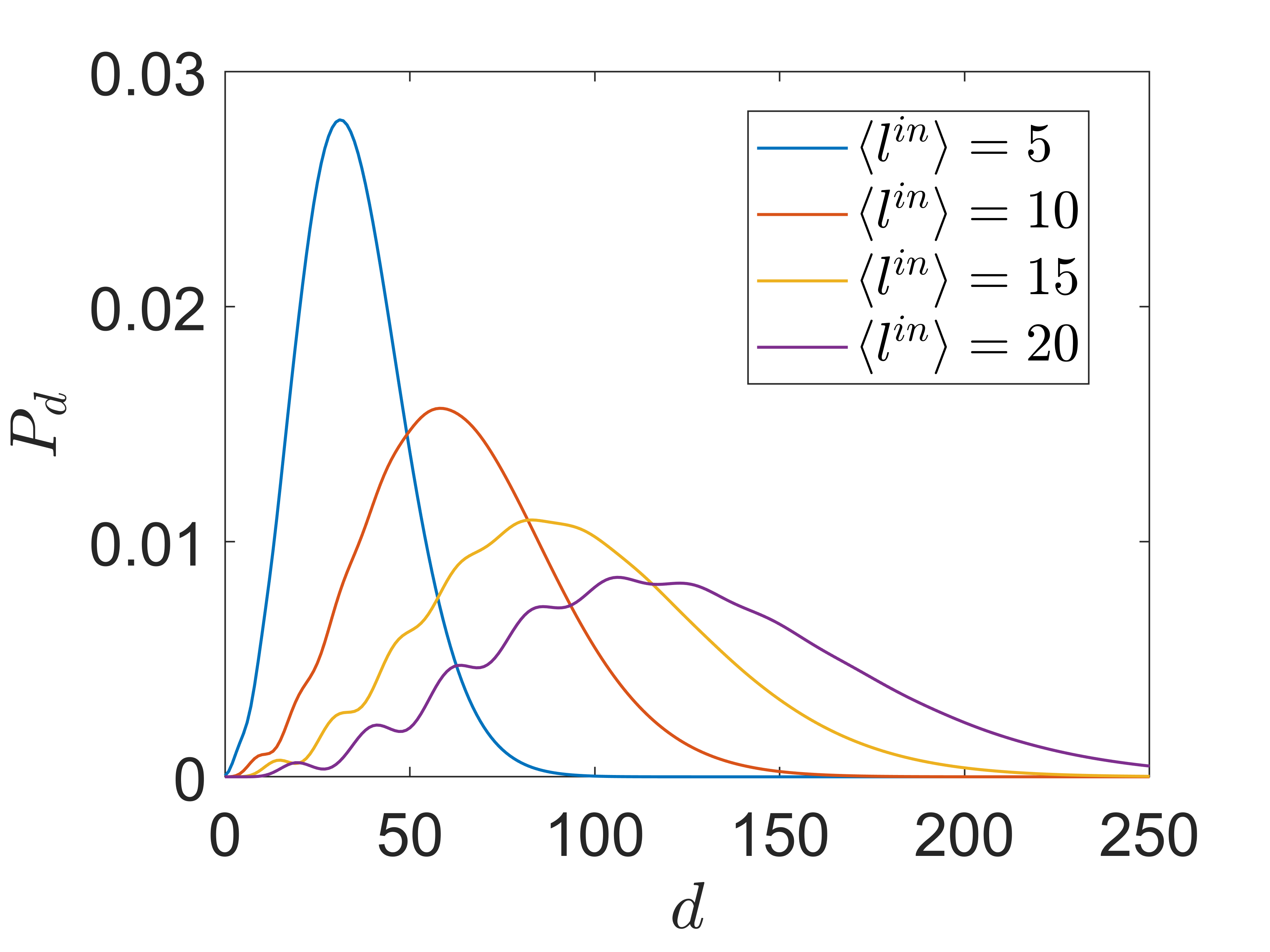}
	\caption{
		{\bf Both inner and outer networks have Poisson distance distribution.} Analytical results from Eq.~\eqref{eq: Poisson on Poisson} and \eqref{eq: integral} where $\av{l^{out}}=5$.
	}
	\label{fig: poisson}
\end{figure}

\noindent
Thus, under proper conditions, if 
\begin{equation}
P_l = \frac{\lambda^l}{l!}e^{-\lambda},
\end{equation}
where $\lambda= \av{l}$,
then the generating function is as known
\begin{equation}
G(x) = e^{\lambda(x-1)}.
\label{eq: poisson gf}
\end{equation}
Now, if both inner and outer networks have approximately Poisson distribution, then for $P_d$ it is satisfied according to Eq.~\eqref{eq: poisson gf} and \eqref{eq: composition} that
\begin{equation}
G_d(x) = e^{\lambda_{in}(x-1)}e^{\lambda_{out}\left(x^we^{\lambda_{in}(x-1)}-1\right)},
\label{eq: Poisson on Poisson}
\end{equation}
where $\lambda_{in} = \av{l^{in}}$ and $\lambda_{out} = \av{l^{out}}$.\\
Still, it is difficult to find an explicit formula for the Taylor coefficients, which are $P_d$, in order to find some criteria for the emergence of wavy distribution. However, numerical calculation shows that if $\av{l^{in}}$ is large enough relative to $\av{l^{out}}$, then the waves appear. See Fig. \ref{fig: poisson}.

\subsection{Two modules} \label{sec: two modules app}
\begin{figure}[ht]	
	\centering
	\includegraphics[width=0.95\linewidth]{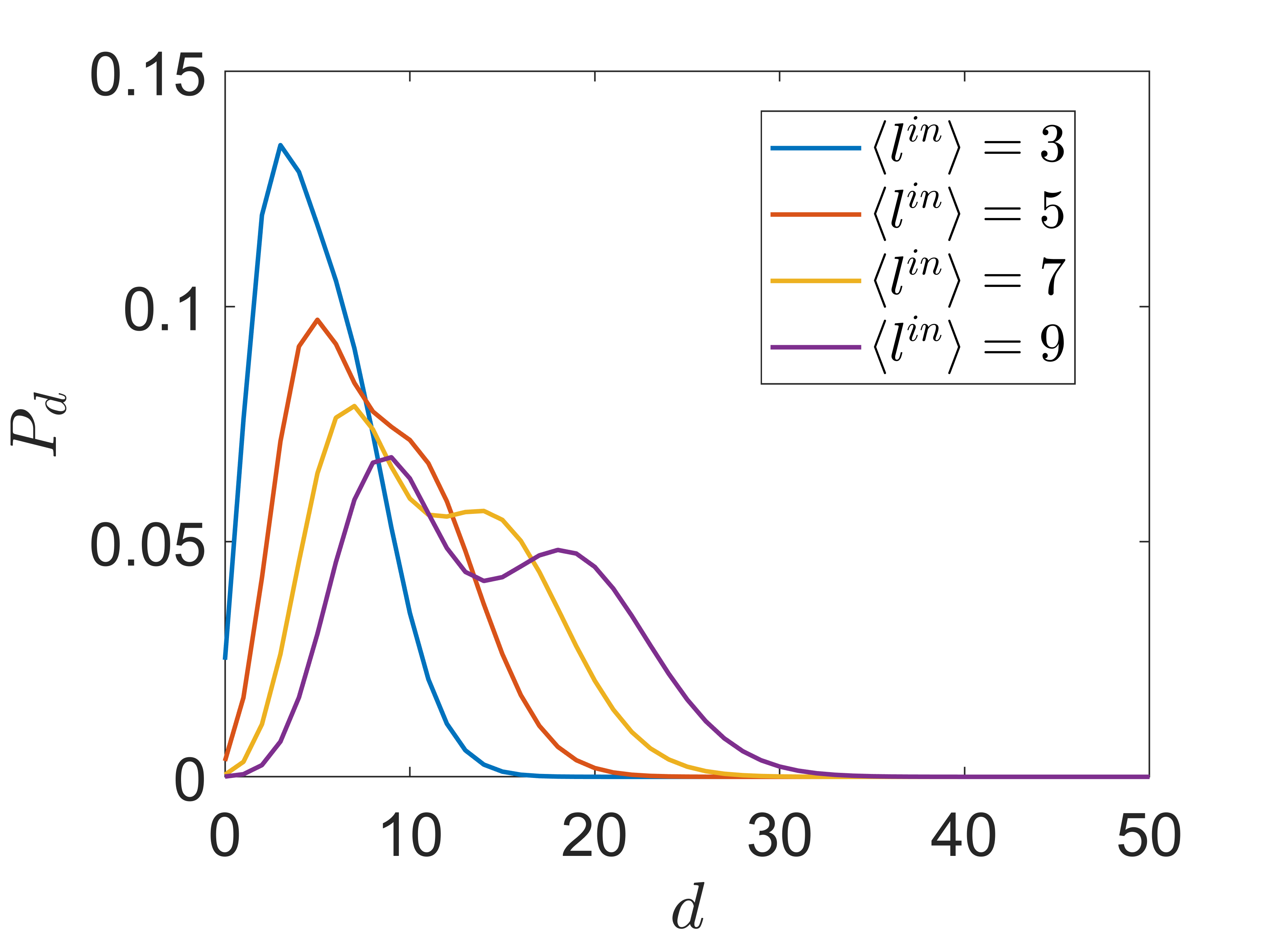}
	\includegraphics[width=0.95\linewidth]{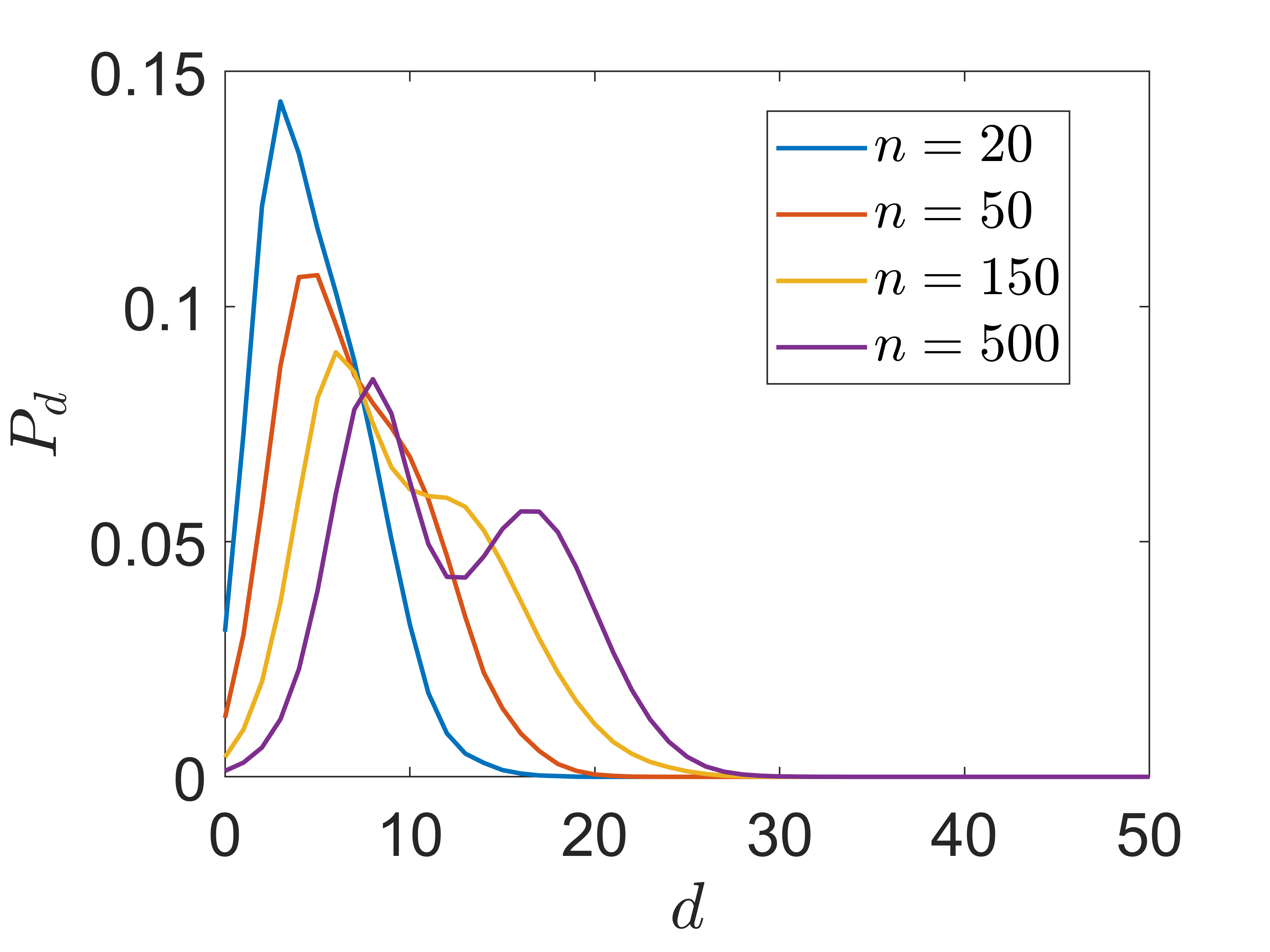}
	\caption{{\bf Emergence of multiple peaks in two modules with Poisson DSPL.} The upper panel is from theory (Eq.~\refeq{eq: Pd poisson of poisson}), and we see that the emergence of two peaks is for $5<\av{l^{in}}<7$ which is in agreement with the value found $\lambda_c=6.27$. In the lower panel we show simulations of ER with $k=2$, and one can see that the transition is slightly above $n=150$, consistent with the finding that approximately $n=160$ gives $\av{l}=6.27$. It works well for $k=2$ because for this range the Poisson approximation holds well as shown in Fig. \ref{fig: ER vs poisson}.
	}
	\label{fig: two poisson}
\end{figure}

To better understand the transition from a single peak to wavy distribution of distances, we study here a simple case which can be fully analyzed analytically. To this end,  we study a network of two connected nodes that satisfies $P_0=1/2,~P_1=1/2$ and $P_l=0$ for any other $l$. Hence, the generating function is
\begin{equation}
G_{\text{two}}(x) = \frac{1}{2}(1+x).
\label{eq: two}
\end{equation}
Let a network of two modules which have a Poisson DSPL, then according to Eqs.~\eqref{eq: two}, \eqref{eq: poisson gf} and \eqref{eq: composition} we obtain 
\begin{equation}
G_{\text{two}\times \text{Poisson}}(x) 
= \frac{1}{2} e^{\lambda(x-1)} \left[1+x^we^{\lambda(x-1)}\right],
\end{equation}
where $\lambda=\av{l^{in}}$. \\
Then we can find the coefficients of the Taylor series
\begin{equation*} 
G_{\text{two}\times \text{Poisson}}(x) 
= \frac{1}{2}\sum_{d=0}^{\infty}\frac{\lambda^d}{d!}e^{-\lambda}x^d + \frac{1}{2}\sum_{d=0}^{\infty}\frac{(2\lambda)^d}{d!}e^{-2\lambda}x^{d+w} ,
\end{equation*}
what yields 
\begin{equation} 
P_d = 
\begin{cases}
\frac{1}{2}\frac{\lambda^d}{d!}e^{-\lambda}, & d<w
\\
\frac{1}{2}\frac{\lambda^d}{d!}e^{-\lambda}
+ \frac{1}{2}\frac{(2\lambda)^{d-w}}{(d-w)!}e^{-2\lambda}, & d\geq w
\end{cases}.
\label{eq: Pd poisson of poisson}
\end{equation}
If $w=1$ then for $d\geq 1$
\begin{equation*} 
P_d = 
\frac{1}{2}\frac{\lambda^d}{d!}e^{-\lambda} \left( 1+\frac{d2^d}{2\lambda} e^{-\lambda} \right).
\end{equation*}
Next we find the ratio
\begin{equation} 
\phi(d,\lambda) \coloneqq \frac{P_{d+1}}{P_d} = 
\frac{\lambda}{d+1} \frac{2\lambda+(d+1)2^{d+1}e^{-\lambda}}{2\lambda+d2^de^{-\lambda}}
\label{eq: ratio}
\end{equation}
because this ratio indicates whether the series $P_d$ increases ($\phi>1$) or decreases ($\phi<1$).
\begin{figure}[ht]	
	\centering
	\includegraphics[width=0.95\linewidth]{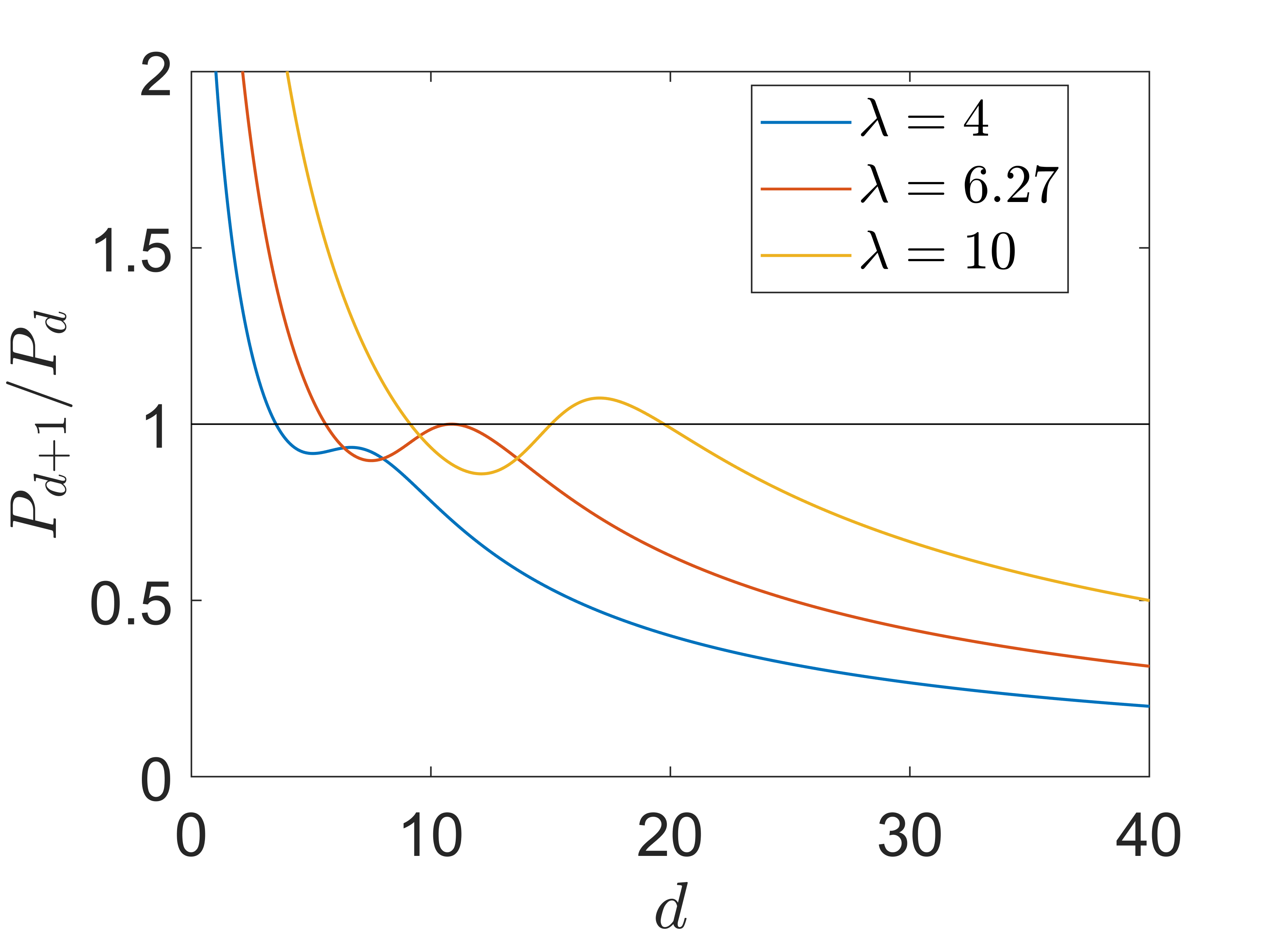}
	\caption{{\bf Analysis of the behavior of  $P_d$ to find out when it increases and decreases for two modules having Poissonian DSPL with average $\lambda$.} The plot presents the ratio $P_{d+1}/P_d$ according to Eq.~\eqref{eq: ratio}. If the ratio is greater than 1 then $P_d$ increases, and if the ratio is lower than 1 then $P_d$ decreases. In this analysis the inter-links weight $w=1$.
	}
	\label{fig: ratio}
\end{figure}
From Fig.~\ref{fig: ratio} one can see that for small $\lambda$ (blue) $P_d$ increases up to some value and then decreases. In contrast, for large $\lambda$ (orange) $P_d$ increases again after decreasing, which indicates a wavy pattern. However, in the transition (red) two conditions are satisfied. 
\begin{equation}
\begin{array}{ll} 
\RN{1} \quad
& \phi(d_c,\lambda_c)  =  1
\\
\RN{2} 
&\frac{\partial \phi}{\partial d} (d_c,\lambda_c)  =  0
\end{array}.
\label{eq: 2eqs criticality}
\end{equation}
Numerical solution of these equations yields $\lambda_c \approx 6.27$. Namely, for two modules which have Poisson DSPL, if $\av{l^{in}}>6.27$, then two peaks will appear. Assuming each module is ER with $k_{in}=2$ (See fig.~\ref{fig: ER vs poisson}), we find numerically that the required size should be approximately $n=160$ in order to satisfy $\av{l}>6.27$. See Fig.~\ref{fig: two poisson} where the simulations results are consistent with this prediction. \\
For different values of $w$ a similar analysis can be done. Higher values of $w$ yield lower values of $\lambda_c$. As example, we find numerically that $\lambda_c\approx3.31$ where $w=2$. \\
In contrast, where $w=0$, then $\lambda_c\approx8.38$.

\subsection{Star graph}
Star graph with $n$ nodes has the following distance distribution
\begin{equation}
N_l = 
\begin{cases}
n, & l=0
\\
2(n-1), & l=1
\\
2\binom{n-1}{2}, & l=2
\end{cases},
\end{equation}
and $P_l=N_l/ n^2$. Then 
\begin{equation}
G_{\text{star}}(x) = P_0+P_1x+P_2x^2.
\end{equation}
$P_1$ is about twice $P_0$, and $P_2$ is about $n$ times $P_0$. Thus, for outer network star graph, if the inner network is such that there are separated peaks, there will be three peaks where the third one is much higher depending on $n$.  

\section{Different cases of inter-links connections}\label{sec: DifferentCasesOfConnectionsTours appendix}
\renewcommand{\theequation}{B.\arabic{equation}}
In this section, we analyze in detail the case of section \ref{sec: DifferentCasesOfConnectionsTours} for which, when entering a module via an interconnected node $i$, we leave this module via different interconnected node $j$ with probability $p$, or, when departing from the module via the same node with probability $1-p$.  See Fig. \ref{fig: illustration lcon}.

We denote $l^{con}$ as the length of the path within the module which was taken during the course, excluding the first and the last modules.
Thus, with probability $1-p$, $l^{con}=0$ (when entering and exiting were via the same node), and  with probability $p$, $l^{con}=l^{in}$ (when entering and exiting the module has been done via different nodes). In the latter case, the distance between the two interconnected nodes is the typical random distance within the module. Therefore, 
\begin{eqnarray}
\begin{aligned}
d &= l_1^{in} + l_2^{in} + wl^{out} + \sum_{i=1}^{l_{out}-1}l_i^{con}
\\
& = l_1^{in} + l_2^{in} + w + \sum_{i=1}^{l_{out}-1} \left( l_i^{con}+w \right),
\end{aligned}
\end{eqnarray}
and if $l^{out}=0$ then $d = l^{in}$.
Hence,
\begin{gather*}
G_d(x) = G_{out}(0)G_{in}(x) + \left[ G_{in}(x) \right]^2 x^w  \times 
\\
\Big(\left[ G_{out}(x)-G_{out}(0) \right] / x \Big) \circ 
\Big( x^w \big( 1-p + pG_{in}(x) \big) \Big).
\end{gather*}
As a result
\begin{equation}
\begin{aligned}
G_d(x) & = G_{out}(0)G_{in}(x) +
\\
&  \left[ G_{in}(x) \right]^2 \frac{  G_{out}\left( x^w \big( 1-p + pG_{in}(x) \big) \right)-G_{out}(0) }{ 1-p + pG_{in}(x) }.
\end{aligned}
\label{eq: p-model appendix}
\end{equation}
One can note that the last equation converges nicely to those of chapters \ref{basicModel} (Eq.~\eqref{eq: composition}) and \ref{sec: one interconnected node} (Eq.~\eqref{eq: same inter}) at the limits $p=1$ and $p=0$ respectively.

\end{document}